\documentclass{ws-procs10x7}
\usepackage{balance}

\newcolumntype{d}[1]{D{.}{.}{#1}}

\def\Journal#1#2#3#4{{\it #1} {\bf #2}, #3 (#4)}

\makeindex
\begin{document}

\title{Microscopic Theory of Current-Spin Interaction \\ in Ferromagnets}

\author{H. Kohno$^*$, S. Kawabata, T. Noguchi, S. Ueta}

\address{Department of Materials Engineering Science, 
Graduate School of Engineering Science, \\
Osaka University, Toyonaka, 
Osaka 560-8531, Japan \\
$^*$E-mail: kohno@mp.es.osaka-u.ac.jp}

\author{J. Shibata}

\address{Kanagawa Institute of Technology, Atsugi, Kanagawa, 243-0292, Japan}

\author{G. Tatara}

\address{Graduate School of Science, Tokyo Metropolitan University, Hachioji, Tokyo 192-0397, Japan}

%%%%%%%%%%%%%%%%%%%%%%%%%%%%%%%%%%%%%%%%%%%%%%%%%%%%%%%%%%%%%%%%%%%%%%%%%
% You may repeat \author \address as often as necessary                 %
%%%%%%%%%%%%%%%%%%%%%%%%%%%%%%%%%%%%%%%%%%%%%%%%%%%%%%%%%%%%%%%%%%%%%%%%%

\twocolumn[\maketitle\abstract{
Interplay between magnetization dynamics and electric current in a conducting 
ferromagnet is theoretically studied based on a microscopic model calculation. 
 First, the effects of the current on magnetization dynamics (spin torques) are studied with special attention to the \lq\lq dissipative" torques arising from spin-relaxation processes of conduction electrons. 
 Next, an analysis is given of the \lq\lq spin motive force", namely, 
a spin-dependent \lq voltage' generation due to magnetization dynamics, 
which is the reaction to spin torques. 
 Finally, an attempt is presented of a unified description of these effects.}
]

\section{Introduction}

The fact that electrons have spin degree of freedom 
as well as electric charge enables us to control, in principle, 
magnetism by electrical means, and vice versa, 
without recourse to the relativistic effect of spin-orbit coupling. 
 This type of magnetoelectric coupling 
has been actively studied over these two decades 
based on nanostructured ferromagnets, where the interplay of electric 
current and magnetization leads to 
giant/tunnel magnetoresistance, current-induced magnetization reversal, 
and so on.\cite{Maekawa06,Xu07}

Microscopic origin of such phenomena is the $s$-$d$ exchange interaction 
\begin{equation}
 H_{sd}  
= - M \int d^3 x \, {\bm n}(x) \!\cdot\! \hat {\bm \sigma} (x) ,
\label{eq:H_sd}
\end{equation}
between the spin $\hat {\bm \sigma}(x)$ of conduction electrons and 
magnetization ${\bm n}(x)$. 
 For example, if an electron moves through a magnetization texture 
${\bm n}(x)$, 
its spin feels a time-dependent \lq field' $M{\bm n}$ and is affected. 
 The electron, in turn, exerts a reaction 
torque\cite{Berger84,Berger92}\,(spin torque)
\begin{equation}
 {\bm t}_{sd} = M {\bm n}(x) \times \langle \hat {\bm \sigma} (x) \rangle ,
\label{eq:t_sd_0}
\end{equation}
on the magnetization, 
which enables us to control the magnetization by current.

In this paper, we present our microsopic study on the spin torque 
and its reciprocal effect (spin motive force). 
 The magnetization is treated as a classical object, whereas 
electrons are treated quantum-mechanically. 

\section{Spin torques}

\subsection{Case of domain wall}

 To illustrate how an electric current flowing in a ferromagnet affects 
the magnetization dynamics, let us first consider a magnetic domain wall 
(DW) as an example.\cite{Berger84,Berger92,Yamaguchi04,TK04}

\begin{figure}[t]
  \begin{center}
  \includegraphics[scale=0.4]{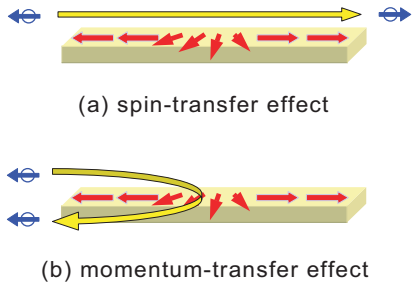}
  \vskip 0mm
  \end{center}
\caption{
Two effects of electric current on a domain wall (DW) via the $s$-$d$ 
exchange interaction.
(a) Adiabatically transmitted electron transfers spin angular momentum 
to the DW, and exerts a torque (in a narrow sense) on the DW.
(b) Reflected electron transfers linear momentum to the DW, and exerts a force 
on the DW. }
\label{fig:DW}
\end{figure}

 For a rigid DW, there are two distinct effects of the current. 
 If a conduction electron passes through the DW adiabatically and 
its spin is flipped after the passage (Fig.\ref{fig:DW}(a)), 
this change of electron spin should be compensated by the change of 
magnetization owing to total angular momentum conservation, 
thereby driving the DW. 
 This is the celebrated spin-transfer effect. 
 If, instead, an electron is reflected by the DW, a linear momentum is 
transferred to the DW and the electron exerts a force 
on it (Fig.\ref{fig:DW}(b)).

 The latter process is nonadiabatic, 
and will be negligible for a \lq thick' DW as realized in typical 
metallic magnets. 
 However, if the electron system admits spin-relaxation processes, 
a new {\it adiabatic} torque (called $\beta$-term, see below) 
arises which has the same effect 
({\it i.e.}, force) on a DW and crucially affects the 
dynamics of the DW.\cite{Zhang04,Thiaville05,Barnes05,TKS08}

\subsection{Landau-Lifshitz-Gilbert equation under current}

 For general but slowly-varying (in space and time) magnetization 
configurations, the dynamics is described by the Landau-Lifshitz-Gilbert (LLG) 
equation, 
\begin{equation}
  \dot {\bm n} = \gamma_0 {\bm H}_{\rm eff} \times {\bm n} 
  + \alpha_0 \dot {\bm n} \times {\bm n} + {\bm t}_{sd}' .
\label{eq:LLG}
\end{equation}
 Here, ${\bm n} = {\bm n}({\bm r},t)$ is a unit-vector field representing the 
$d$-spin direction, and the dot represents time derivative. 
 The first term, a precessional torque around the effective field 
$\gamma_0 {\bm H}_{\rm eff}$, 
and the second term (Gilbert damping) come from processes 
without conduction electrons. 
 The effects of conduction electrons are contained in the third term, 
${\bm t}_{sd}' \equiv  {\bm t}_{sd}$  $\times (a^3/\hbar S)$ 
($S$: magnitude of $d$ spin, $a^3$: volume per $d$-spin), 
called spin torque.

In this paper, we focus on {\it adiabatic spin torques},\cite{com1} 
which are first order in space/time derivative and are expressed as 
\begin{eqnarray}
 {\bm t}_{sd}^{\prime} 
&=& - ({\bm v}_{\rm s}^0 \!\cdot\! {\bm \nabla})\,  {\bm n} 
  - \beta_{\rm sr} \, {\bm n} \times ({\bm v}_{\rm s}^0 \cdot\! {\bm \nabla}) 
    \, {\bm n} 
\nonumber \\
&{}& - \alpha_{\rm sr} \, ({\bm n} \times \dot {\bm n}) 
  - \frac{\delta S}{S} \, \dot {\bm n} . 
\label{eq:t_sd_1}
\end{eqnarray}
 The first term on the right-hand side is the celebrated 
spin-transfer torque,\cite{BJZ98} where 
\begin{equation}
 {\bm v}_{\rm s}^0 = - \frac{a^3}{2eS} \, {\bm j}_{\rm s} 
\label{eq:vs0}
\end{equation}
is the (unrenormalized) \lq\lq spin-transfer velocity'', 
with ${\bm j}_{\rm s} = {\bm j}_\uparrow - {\bm j}_\uparrow$ 
being the spin-current density. 
 The second term, called \lq $\beta$-term',\cite{Thiaville05} 
comes from spin-relaxation processes of electrons,\cite{Zhang04} 
and acts as a force on a rigid DW. 
 Here $\beta_{\rm sr}$ is a dimensionless constant. 
 The third term is the Gilbert damping, also resulting from 
spin relaxation of electrons.

 The fourth term contributes as a \lq\lq renormalization'' of 
spin;\cite{Zhang04} 
it can be combined with the term on the left-hand side of eq.(\ref{eq:LLG}) 
to form 
$(1 + \delta S/S) \,\dot {\bm n}= (S_{\rm tot}/S) \, \dot {\bm n}$, 
where 
\begin{eqnarray}
 S_{\rm tot} = S + \delta S ,
\label{eq:S_tot}
\end{eqnarray}
is the total (\lq\lq renormalized'') spin 
with $\delta S$ being the contribution from conduction electrons. 
 Then, dividing both sides of the equation by $S_{\rm tot}/S$, 
we arrive at 
\begin{eqnarray}
  \dot {\bm n} 
&=& \gamma {\bm H}_{\rm eff} \times {\bm n} 
- \alpha \, ({\bm n} \times \dot {\bm n}) 
\nonumber \\
&{}& - ({\bm v}_{\rm s} \!\cdot\! {\bm \nabla})\,  {\bm n} 
     - \beta \, {\bm n} \times 
      ({\bm v}_{\rm s} \cdot\! {\bm \nabla}) \, {\bm n} ,
\label{eq:LLG2}
\end{eqnarray}
where 
$\gamma = (S/S_{\rm tot}) \, \gamma_0$, 
$\alpha = (S/S_{\rm tot}) (\alpha_0 + \alpha_{\rm sr} )$, 
$\beta = \beta_{\rm sr}$, 
and 
\begin{eqnarray}
  {\bm v}_{\rm s}
= \frac{S}{S_{\rm tot}} \, {\bm v}_{\rm s}^0 
= - \frac{a^3}{2eS_{\rm tot}} {\bm j}_{\rm s} ,
\label{eq:vs}
\end{eqnarray}
is the \lq\lq renormalized'' spin-transfer velocity. 
 Note that $\beta$ is not renormalized by this procedure.

 In the parameter space of the LLG equation (\ref{eq:LLG2}), 
the manifold of 
$\alpha = \beta$ provides a very special case for the dynamics. 
 For example, any static solution ${\bm n}({\bm r})$ 
in the absence of spin current can be used to construct a solution 
${\bm n}({\bm r}- {\bm v}_{\rm s} t)$ in the presence of spin current 
${\bm v}_{\rm s}$ if $\alpha = \beta$. 
 Since the controversy on the current-driven domain-wall 
motion,\cite{TK04,Barnes05} 
whether the relation $\alpha = \beta$ holds generally 
or not has been a theoretical issue.

 The relation $\alpha = \beta$ was originally suggested in 
ref.\cite{Barnes05} based on the assumption of Galilean invariance 
of the system. 
 Although one may argue that the Galilean invariance should be valid 
for long-wavelength and low-frequency dynamics in which the 
underlying lattice structure is irrelevant, the $\alpha$ and $\beta$ 
come from spin-relaxation processes,\cite{Zhang04} 
which are usually intimately related to the lattice, 
{\it e.g.}, through the spin-orbit coupling. 
 Also, for a many-electron system having Fermi surfaces, Galilean 
invariance is not an obvious property. 
 Therefore, it is desired to carry out a fully microscopic calculation 
without introducing any phenomenological assumptions 
once a microscopic model is fixed.

\subsection{Microscopic model}

 For conceptual simplicity, we take a localized picture for ferromagnetism, 
and consider the so-called $s$-$d$ model  
consisting of localized $d$ spins, ${\bm S}=S{\bm n}$, and conducting $s$ 
electrons (as we already used in the previous sections). 
 They are mutually coupled {\it via} the $s$-$d$ exchange interaction 
$H_{sd}$ [eq.(\ref{eq:H_sd})] and obey, respectively, the LLG equation 
(\ref{eq:LLG}) and the Schr\"odinger equation 
\begin{equation}
 i \hbar \dot c 
= \left[ - \frac{\hbar^2}{2m} \nabla^2 
   - M {\bm n} \!\cdot\! {\bm \sigma} + V_{\rm imp}  \right] c . 
\label{eq:Sch}
\end{equation}
 The impurity potential $V_{\rm imp}$ includes potential scattering 
as well as spin scattering 
\begin{equation}
 V_{\rm imp}^{\rm s} 
=  u_{\rm s} \sum_j {\bm S_j} \!\cdot\! {\bm \sigma} \, 
             \delta ({\bm r} - {\bm R}_j')  
\label{eq:Vimp}
\end{equation}
due to quenched magnetic impurities ${\bm S_j}$. 
 The latter has been introduced as a microscopic modeling of 
spin-relaxation processes. 
 The averaging over the impurity spin direction is taken as 
$\overline{S_i^\alpha} = 0$ and 
\begin{equation}
  \overline{S_i^\alpha S_j^\beta} 
= \frac{1}{3} S_{\rm imp}^2 \delta_{ij} \delta^{\alpha\beta} .
\label{eq:SS}
\end{equation}

 To obtain the torque ${\bm t}_{sd}$, 
we calculate the $s$-electron spin density 
$\langle \hat {\bm \sigma}_\perp \rangle_{\rm ne}$ 
[see eq.(\ref{eq:t_sd_0})]. 
($\perp$ means perpendicular component to ${\bm n}$.) 
 Here the average $\langle \cdots \rangle_{\rm ne}$ 
is taken in the following nonequilibrium states for electrons 
depending on the type of the torque.

(a) Nonequilibrium states under the influence of uniform but 
{\it time-dependent magnetization}. 
  This leads to torques with time derivative of ${\bm n}$,
namely, Gilbert damping and spin renormalization. 

(b) Nonequilibrium states with {\it current flow} under static but 
{\it spatially-varying magnetization}.  
 This leads to current-induced torques, namely, 
spin-transfer torque and the $\beta$-term.

\subsection{Small-amplitude method} 

 In the presence of spin rotational symmetry in the electron system 
(except for $H_{sd}$), 
adiabatic spin torques are expressed as
\begin{equation}
 {\bm t}_{sd}
= a_\mu \partial_\mu {\bm n} 
  + b_\mu \, ({\bm n} \times \partial_\mu {\bm n}) ,
\label{eq:t_sd_2} 
\end{equation}
where $a_\mu$ and $b_\mu$ are the coefficients,  
and summing over $\mu =1,2,3$ (space components) and 0 (time) 
is understood. 
 The corresponding $s$-electron spin polarization is given by 
\begin{equation}
\langle {\bm \sigma}_\perp \rangle_{\rm ne} 
= \frac{1}{M} \left[\, 
    b_\mu \partial_\mu {\bm n} 
  - a_\mu \, ({\bm n} \times \partial_\mu {\bm n})  \,\right] . 
\label{eq:sigma_perp} 
\end{equation}
 The coefficients, $a_\mu$ and $b_\mu$, can be determined by 
considering small transverse fluctuations, 
${\bm u} = (u^x,u^y,0)$, $|{\bm u}| \ll 1$, 
around a uniformly magnetized state, ${\bm n} = \hat z \equiv (0,0,1)$, 
such that ${\bm n} = \hat z + {\bm u} + {\cal O}(u^2)$, 
and retain the terms first order in ${\bm u}$ as \cite{TSBB06,KTS06,Duine07} 
\begin{equation}
\langle {\bm \sigma}_\perp \rangle_{\rm ne} 
= \frac{1}{M} \left[\, 
b_\mu \partial_\mu {\bm u} 
   - a_\mu (\hat z \times \partial_\mu {\bm u}) \,\right] .
\label{eq:sigma_u}
\end{equation}
 Then $a_\mu$ and $b_\mu$ are given as 
linear-response coefficients, which are evaluated in the uniformly 
magnetized state, ${\bm u}={\bm 0}$.

\begin{figure}[t]
  \begin{center}
  \includegraphics[scale=0.4]{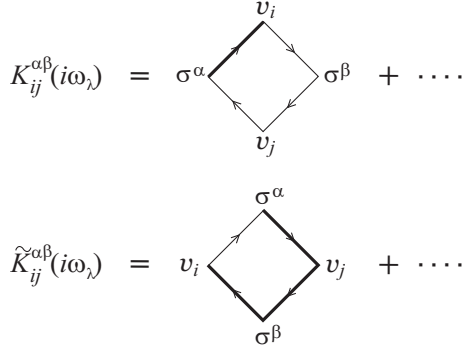}
  \vskip 0mm
  \end{center}
\caption{Diagrammatic expression for spin torque (upper panel) 
and spin motive force (lower panel). 
 The thick lines carry the external frequency $i\omega_\lambda$.}
\label{fig:diagram}
\end{figure}

 To calculate current-induced torques for example, we assume 
a static configuration, 
${\bm n}({\bm r}) = \hat z + {\bm u}({\bm r})$, and 
introduce a d.c. electric field ${\bm E}$ to produce a current-carrying 
state. 
 We calculate ${\bm \sigma}_\perp$ by first applying the 
linear-response theory to extract ${\bm E}$ as 
\begin{equation}
 \langle \hat\sigma^\alpha_\perp ({\bm q}) \rangle_{\rm ne} 
  =  \lim_{\omega \to 0}
    \frac{ K_i^\alpha ({\bm q},\omega +i0)}{i\omega} \, E_i .
\end{equation}
 The linear-response coefficient 
\begin{equation}
 K_i^\alpha ({\bm q}, i\omega_\lambda ) 
 = \int_0^\beta d\tau \, {\rm e}^{i\omega_\lambda \tau} \, 
    \langle \, {\rm T}_\tau \, \hat\sigma_\perp^\alpha ({\bm q}, \tau) 
            \, J_i \, \rangle   \ \ \ 
\label{eq:K}
\end{equation} 
is the correlation function of spin $\hat\sigma$ and electric current 
${\bm J}$, 
which can be non-vanishing in the presence of non-uniform spin texture 
${\bm u}({\bm r}) = u_{\bm q} {\rm e}^{i{\bm q}\cdot{\bm r}}$. 
 Extracting $u^\beta$ and $q_j$ as 
\begin{equation}
  K_i^{\alpha} ({\bm q}, i\omega_\lambda ) 
= - eM K_{ij}^{\alpha\beta} (i\omega_\lambda ) q_j u^\beta_{\bm q}, 
\end{equation}
we have calculated the coefficient $K_{ij}^{\alpha\beta}$, 
which is expressed by the upper diagram in Fig. \ref{fig:diagram}.

 The results are given, in the lowest non-trivial order in the electron 
damping, by \cite{KTS06,Duine07}
\begin{eqnarray}
 \delta S &=& \frac{1}{2} \, \rho_{\rm s} a^3 , 
\label{eq:result_deltaS}
\\
  {\bm v}_{\rm s} 
&=& - \frac{a^3}{2e\, (S+\delta S)} \, {\bm j}_{\rm s} , 
\label{eq:result_vs}
\\
  \alpha 
&=&  \frac{a^3 \nu_+}{4(S+\delta S)} \!\cdot\! \frac{\hbar}{\tau_{\rm s}} 
   + \frac{S}{S+\delta S} \, \alpha_0 , 
\label{eq:result_alpha}
\\
  \beta 
&=&  \frac{\hbar}{2M\tau_{\rm s}} .
\label{eq:result_beta}
\end{eqnarray}
 Here $\rho_{\rm s}  = n_\uparrow - n_\downarrow$ is the $s$-electron 
spin density, 
$\nu_\pm = \nu_\uparrow \pm \nu_\downarrow$ is the density of states, and 
${\bm j}_{\rm s} = \sigma_{\rm s}{\bm E}
 = {\bm j}_\uparrow - {\bm j}_\downarrow $ is the spin current, 
with 
$\sigma_{\rm s} = 
 \sigma_\uparrow - \sigma_\downarrow$ 
being the \lq\lq spin conductivity''. 
($\sigma_{\uparrow (\downarrow)}$ is the conductivity of 
majority- (minority-) spin electrons.)
 We have defined the spin-relaxation time $\tau_{\rm s}$ by 
\begin{equation}
  \frac{\hbar}{\tau_{\rm s}} 
 =  \frac{4\pi}{3} \, n_{\rm s}u_{\rm s}^2 \, S_{\rm imp}^2 \, \nu_+ . 
\label{eq:tau_s}
\end{equation}
 As expected, only the spin scattering ($\sim \tau_{\rm s}^{-1}$) 
contributes to $\alpha$ and $\beta$, 
and the potential scattering does not. 

 The ratio $\beta / \alpha$ cannot be unity in general 
for the two-component $s$-$d$ model, since it contains mutually independent 
quantities, 
{\it e.g.}, $S$ of $d$ electrons and $\delta S$ of $s$ electrons. 
 For a single-band itinerant ferromagnet, where $\delta S$ gives the 
total moment, the results are obtained by simply putting 
$S=0$ and $\alpha_0 = 0$ in 
eqs.(\ref{eq:result_deltaS})-(\ref{eq:result_beta}). 
 We still see that $\alpha \ne \beta$, but  
it was pointed out that the ratio 
\begin{equation}
  \frac{\beta}{\alpha} 
=  \frac{\rho_{\rm s}}{M \nu_+} 
\simeq 1 + \frac{1}{12} \left(\frac{M}{\varepsilon_{\rm F}} \right)^2 
\end{equation}
is very close to unity.\cite{TSBB06} 
 Even so, if we generalize eq.(\ref{eq:SS}) to the anisotropic one, 
\begin{eqnarray}
  \overline{S_i^\alpha S_j^\beta} 
&=& \delta_{ij} \delta_{\alpha\beta} \times \left\{ \begin{array}{cc} 
    \overline{S_\perp^2} & (\alpha, \beta = x,y) \\ 
    \overline{S_z^2}     & (\alpha, \beta = z) 
    \end{array} \right.
\end{eqnarray}
we have 
\begin{equation}
  \frac{\beta}{\alpha} 
=  \frac{3 \overline{S_\perp^2} + \overline{S_z^2} }
        {\, 2 \, (\overline{S_\perp^2} + \overline{S_z^2}) \,} , 
\end{equation}
which ranges from $1/2$ (for $\overline{S_\perp^2} \ll \overline{S_z^2}$) 
to $3/2$ (for $\overline{S_\perp^2} \gg \overline{S_z^2}$). 
 Therefore, we conclude that $\alpha \ne \beta$ in general, and 
that the value $\beta / \alpha$ is very 
sensitive to the details of the spin-relaxation mechanism.

 The \lq\lq $\beta$-term'' due to spin relaxation was first derived 
by Zhang and Li 
 based on a phenomenological spin-diffusion equation.\cite{Zhang04} 
 Their results can be written as 
\begin{equation}
  \alpha_{\rm ZL} 
=  \frac{\delta S}{S+\delta S} \!\cdot\! \frac{\hbar}{2 M \tau_{\rm s}} , 
\label{eq:alpha_ZL}
\end{equation}
and $\beta_{\rm ZL} = \hbar /2 M \tau_{\rm s}$, 
thus predict \lq\lq $\alpha = \beta$" for a single-band itinerant 
ferromagnet, $S=0$. 
 So far, all phenomenologial theories predict $\alpha = \beta$, 
in contrast to the present microscopic results\cite{KTS06} 
showing $\alpha \ne \beta$ in general.

\subsection{Gauge-field method}

 The treatment in the previous subsection is based on the assumption of 
rotational symmetry in spin space of electrons; otherwise it is limited to 
small-amplitude magnetization dynamics around a uniformly magnetized state. 
 To treat finite-amplitude dynamics directly, we introduce in this section 
a local/instantaneous spin frame (\lq\lq adiabatic frame'') 
for $s$ electrons.\cite{Korenman77,TF94} 
 In this frame, the spin quantization axis of $s$ electrons is taken 
to be the local/instantaneous $d$-spin direction, ${\bm n}$. 
 The electron spinor $a(x)$ in the new frame is related to the 
original spinor $c(x)$ as $c(x) = U(x) a(x)$, where $U$ is a 
$2\times 2$ unitary matrix satisfying 
$ c^\dagger ({\bm n} \!\cdot\! {\bm \sigma}) c 
 = a^\dagger \sigma^z a$. 
 The $a$-electrons then obey the equation,
\begin{eqnarray}
&{}& 
 i\hbar \left( \frac{\partial}{\partial t} 
 + i A_{_0} \right) a(x) 
\\
&{}& = \left[ - \frac{\hbar^{2}}{2m}  \left(\nabla_i + iA_i \right)^2 
- M \sigma_z + \tilde V_{\rm imp} \, \right] a(x) , 
\nonumber 
\label{eq:Sch_a}
\end{eqnarray}
which is characterized by a constant magnetization $M \sigma_z$ 
and an SU(2) gauge field 
\begin{equation}
 A_\mu = -i U^\dagger (\partial_\mu U) = A^\alpha_\mu \sigma^\alpha 
\equiv {\bm A}_\mu \!\cdot {\bm \sigma} . 
\label{eq:A_UdU}
\end{equation}
 This gauge field expresses the influence of temporal ($\mu=0$) or spatial 
($\mu=1,2,3$) variation of ${\bm n}$.

 The adiabatic torques in eq.(\ref{eq:t_sd_2}) 
follow from the following expression\cite{KS07} 
\begin{equation}
 \langle \tilde {\bm \sigma}_\perp \rangle_{\rm ne} 
= \frac{2}{M} \left[ 
   a_\mu {\bm A}^\perp_\mu 
  + b_\mu (\hat z \times {\bm A}^\perp_\mu )  \right] ,
\label{eq:sigma_perp_A}
\end{equation}
obtained in the first order in ${\bm A}_\mu$. 
 Here 
$\langle \tilde {\bm \sigma} \rangle 
 \equiv \langle a^\dagger {\bm \sigma} a \rangle$ 
is the electron spin density in the adiabatic frame, 
and 
$\tilde {\bm \sigma}_\perp$ and $ {\bm A}^\perp_\mu$ 
are those projected onto the $xy$-plane. 
 The coefficients $a_\mu$ and $b_\mu$ can be calculated as 
linear-response coefficients. 
 The results for $\delta S, {\bm v}_{\rm s}$ and $\beta$ 
thus obtained coincide with eqs.(\ref{eq:result_deltaS}), 
(\ref{eq:result_vs}), and (\ref{eq:result_beta}). 
 However, it leads to $\alpha_{\rm sr} = 0$ and 
fails to produce the Gilbert damping.

\begin{figure}[t]
  \begin{center}
  \includegraphics[scale=0.3]{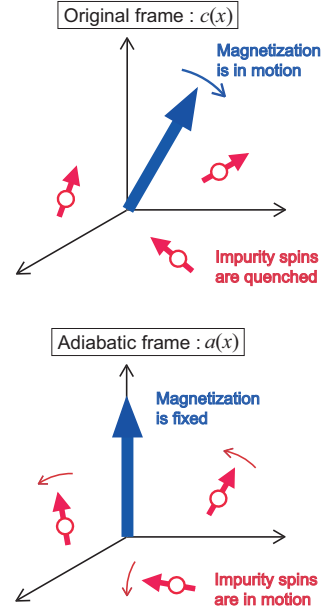}
  \vskip 0mm
  \end{center}
\caption{Upper panel (lower panel) shows magnetization vector 
${\bm n}(t)$ ($\hat z$) and 
impurity spins ${\bm S}_j$ ($\tilde {\bm S}_j (t)$) 
in the original frame (adiabatic frame).}
\label{fig:Gilbert}
\end{figure}

 This difficulty has been resolved\cite{KS07} 
by noting 
that the impurity spins, which are static (quenched) in the original frame, 
become time-dependent in the adiabatic frame: 
\begin{equation}
 \tilde {\bm S_j}(t) = {}^t{\cal R}(t) {\bm S_j} .
\label{eq:SS_RS}
\end{equation}
(See Fig.~\ref{fig:Gilbert}.) 
 Here ${\cal R}$ is a $3\times 3$ orthogonal matrix 
representing the same rotation as $U$ but acting on three-component vectors. 
 From the time dependence of $\tilde {\bm S_j}(t)$ or ${\cal R}(t)$, 
the SU(2) gauge field can arise as 
\begin{equation}
 [ {\cal R}(t) \,^t \dot {\cal R}(t) ] ^{\alpha \beta} 
= 2 \varepsilon^{\alpha \beta \gamma} A_0^\gamma (t). 
\label{eq:RdR_A0}
\end{equation}
 In fact, explicit evaluation of 
$\langle \tilde {\bm \sigma}_\perp \rangle_{\rm ne}$ 
in second order in $ \tilde {\bm S_j}(t)$ (nonlinear response) gives 
\begin{equation}
 \langle \tilde {\bm \sigma}_\perp \rangle_{\rm ne} 
= - \frac{2\pi\hbar}{3M} n_{\rm s} u_{\rm s}^2 S_{\rm imp}^2 
  \nu_+^2 \, (\hat z \times {\bm A}_0^\perp )  , 
\label{eq:sigma_gauge}
\end{equation}
leading to the Gilbert damping which coincides with the first 
term of eq.(\ref{eq:result_alpha}).

 The above calculation provides us a new picture of Gilbert damping; 
while the spins of $s$ electrons tend to follow ${\bm n}(t)$, 
it is at the same time pinned by the quenched impurity spins, 
and this frustration gives rise to the Gilbert damping. 
 This picture also applies to the case where 
spin relaxation originates from spin-orbit coupling.\cite{Kawabata}

\section{Spin motive force}

 As a reaction to spin torques, magnetization dynamics in turn exerts a 
spin-dependent force, called spin motive force, on electrons.\cite{Berger86,Volovik,Stern92,Barnes06a,Duine07b,Stamenova07,Yang07,Tserkovnyak07,Yang08,Brataas02} 
 According to Stern,\cite{Stern92} this effect arises from the 
time-dependent spin Berry phase, which we interpret in our context as 
arising as a combined effect of temporal variation and spatial variation 
of magnetization. 
 Here we present a simple argument using the results obtained in 
the previous section.

 We apply the small-amplitude method, and consider 
a small fluctuation of the form, 
\begin{equation}
  {\bm u} ({\bm r},t)
 = {\bm u}_1 \, {\rm e}^{-i\omega t} 
 + {\bm u}_2 \, {\rm e}^{i{\bm q}\cdot{\bm r}} , 
\label{eq:u1u2}
\end{equation}
to calculate the current density in the first order in 
$\dot {\bm u}$ and $\nabla {\bm u}$, {\it i.e.}, 
in $\omega {\bm u}_1$ and $q_j {\bm u}_2$: 
\begin{equation}
  \langle j_i ({\bm q}) 
  \rangle_{_{\rm ne}} 
 = -eM 
   \frac{ \, \tilde K_{ij}^{\alpha\beta} (\omega ) \,\, }{i\omega}  \, 
\omega \, u_1^\alpha \!\cdot q_j u_2^\beta . 
\end{equation}
 The coefficient $\tilde K_{ij}^{\alpha\beta}$ (see Fig.\ref{fig:diagram}) 
can be shown to 
be related to $K_{ij}^{\alpha\beta}$ of the spin torque as 
\begin{equation}
  K_{ij}^{\alpha\beta} (i\omega_\lambda )  
= \tilde K_{ij}^{\alpha\beta} (-i\omega_\lambda ) . 
\end{equation}
 Therefore, using the results of \S 2-4, we readily obtain 
${\bm j} = \sigma_{\rm s} {\bm E}_{\rm s}$, 
where 
\begin{eqnarray} 
 E_{{\rm s}, i} = \, \frac{\hbar}{2e} \, 
 \Bigl[\, {\bm n} \!\cdot\! (\partial_i {\bm n} \times \dot {\bm n})
  + \beta \, (\dot {\bm n} \!\cdot\! \partial_i {\bm n}) \,\Bigr] . \ 
\end{eqnarray} 
 From 
${\bm j} = \sigma_\uparrow {\bm E}_{\rm s} 
 + \sigma_\downarrow (-{\bm E}_{\rm s})$, 
we may identify ${\bm E}_{\rm s}$ to be a spin-dependent \lq electric' 
field, or $-e{\bm E}_{\rm s}$ to be the spin motive force, 
in the sense that majority- (minority-) spin electrons feel an effective 
\lq electric' field of ${\bm E}_{\rm s}$ ($-{\bm E}_{\rm s}$). 
 The second term, containing the same $\beta$ parameter as the spin torque, 
is due to spin relaxation, and was first reported by Duine.\cite{Duine07b} 

 More general calculation without assuming the form of 
eq.(\ref{eq:u1u2}),\cite{Noguchi} 
as well as the one based on the gauge-field method\cite{Shibata} 
will be reported elsewhere.

\section{Effective gauge-field action}

Spin torque and spin motive force are action and reaction to each other, 
and should be derived from the same term in the effective action. 
 This kind of study has been done by Duine {\it et al.}\cite{Duine07} 
based on the real-time, small-amplitude formalism. 
 Here we present a treatment based on the imaginary-time, 
gauge-field formalism. 
 It should be noted that dynamical/dissipative processes can also 
be treated with imaginary time.

 We introduce an electromagnetic vector potential ${\bm A}^{\rm em}$ 
to drive the non-equilibrium Ohmic current in a ferromagnet, 
and eliminate the $a$-electrons. 
 Up to the second order in ${\bm A}^{\rm em}$ and the SU(2) gauge field 
${\bm A}^\alpha$, the effective action ${\cal S}$ is obtained as\cite{Ueta} 
\begin{eqnarray}
 {\cal S} 
&=&  \int_0^{\beta'} d\tau \int \frac{d{\bm r}}{a^3} 
   \left[ 2i \hbar S_{\rm tot} A_0^z 
   + \frac{J_{\rm eff}}{2} (\partial_i {\bm n})^2  \right] 
\nonumber \\
&+& \int_0^{\beta'} d\tau \int_0^{\beta'} d\tau' \int d{\bm r} 
    I(\tau -\tau') 
\nonumber \\
&\times& \Biggl\{ \left[ 
  \sigma_{\rm s} \frac{\hbar}{e} {\bm A}^z (\tau) 
 + \frac{\sigma_{\rm c}}{2} {\bm A}^{\rm em} (\tau) \right] 
  \!\cdot\! {\bm A}^{\rm em} (\tau') 
\nonumber \\
&{}& \ \ \ + \, c_{\alpha\beta} \, 
  [{\cal R} (\tau) {}^t{\cal R}(\tau')]^{\alpha\beta} 
  \Biggr\} . 
\end{eqnarray}
 Here $\beta' \equiv (k_{\rm B}T)^{-1} = \infty$ is the inverse temperature, 
$J_{\rm eff} = J_{dd} S^2 + J_{ss} (\delta S)^2 $,\cite{TF94} 
and 
\begin{equation}
 c_{\alpha\beta} 
= \frac{\pi}{6} n_{\rm s} u_{\rm s}^2 S_{\rm imp}^2 
  \left[ 2 \nu_\uparrow \nu_\downarrow \delta^{\alpha\beta} 
      + \nu_-^2 \delta^{\alpha z} \delta^{\beta z} 
  \right] . 
\end{equation}
 The kernel $I(\tau -\tau') = - [\pi (\tau -\tau')^2]^{-1}$ 
describes dissipative processes characterized by Ohmic damping, 
as is familiar since the work by Caldeira and Leggett\cite{CL} on macroscopic 
quantum tunneling. 
 The coupling ${\bm A}^z \!\cdot\! {\bm A}^{\rm em}$ describes 
the spin-transfer torque and spin motive force.  
 The term containing ${\cal R} (\tau) {}^t{\cal R}(\tau')$ 
describes Gilbert damping. 

 In fact, by taking the variation of ${\cal S}$ with respect to ${\bm n}$, 
and perform an analytic continuation, $\tau \to it$, 
we obtain the LLG equation consistent with eqs.(\ref{eq:LLG2}), 
(\ref{eq:result_deltaS})-(\ref{eq:result_alpha}) 
but with $\beta = 0$. 
 Similarly, the electric current density is obtained from 
${\bm j} = - \delta {\cal S}/ \delta {\bm A}^{\rm em}$ as 
\begin{equation}  
 {\bm j} 
 = - \frac{\hbar}{e} \sigma_{\rm s} \dot {\bm A}^z 
   - \sigma_{\rm c} \dot {\bm A}^{\rm em} ,
\end{equation} 
from which we can read the existence of the spin motive force as 
$-e {\bm E}_{\rm s} = \hbar \dot {\bm A}^z  
 = (\hbar/2) \,{\bm n} \!\cdot\! (\dot {\bm n} \times \partial_i {\bm n})$. 
 The effective coupling describing the $\beta$-term remains to be derived.

\section{Summary and remarks}

 We have developed a microscopic theory of spin torques and spin 
motive force, and their unified description. 
 Although the present magnetic impurity model 
may not be quite realistic 
as the origin of spin relaxation, 
we expect the present calculation already captures the 
essential features of the current-spin interaction 
including spin-relaxation effects. 
 For quantitative information such as the value of $\beta/\alpha$,
calculations with realistic spin-relaxation mechanisms 
are necessary.

\section*{Acknowledgments}
 We would like to thank 
G.\,\,Bauer, 
A.\,Brataas, 
R.\,\,Duine, 
H.\,\,Fukuyama, 
A.\,H.\,\,MacDonald, 
S. Maekawa, 
Y.\,\,Nakatani, 
Q.\,\,Niu, 
H.\,\,Ohno, 
T.\,\,Ono, 
E.\,\,Saitoh, 
J.\,\,Sinova, 
M.\,\,Stiles, 
Y. Suzuki, 
A.\,Thiaville and 
Y. Tserkovnyak 
for valuable discussions. 
H.~K. is indebted to 
K. Miyake for his continual encouragement.

\end{document}